\documentclass[conference]{IEEEtran}
\IEEEoverridecommandlockouts
\usepackage{cite}
\usepackage{amsmath,amssymb,amsfonts}
\usepackage{algorithmic}
\usepackage{graphicx}
\usepackage{multirow}
\usepackage{textcomp}
\usepackage{xcolor}
\usepackage{xcolor, colortbl}
\usepackage{tabularx,lipsum,amssymb,amsmath,pifont,graphicx,adjustbox}
\usepackage{lipsum}
\usepackage{verbatim}

\usepackage{url}
\def\BibTeX{{\rm B\kern-.05em{\sc i\kern-.025em b}\kern-.08em
    T\kern-.1667em\lower.7ex\hbox{E}\kern-.125emX}}
\begin{document}

\title{Investigating voiced and unvoiced regions of speech for audio deepfake detection}

\author{\IEEEauthorblockN{Ganesh Sivaraman, Hemlata Tak, Elie Khoury}
\IEEEauthorblockA{Pindrop, Atlanta, USA \\
\{gsivaraman, hemlata.tak, ekhoury\}@pindrop.com}

% \author{\IEEEauthorblockN{Anonymous Authors}}
% \and
% \IEEEauthorblockN{Hemlata Tak}
% \IEEEauthorblockA{\textit{Pindrop}\\
% Atlanta, USA \\
% hemlata.tak@pindrop.com}
% \and
% \IEEEauthorblockN{Elie Khoury}
% \IEEEauthorblockA{\textit{Pindrop}\\
% Atlanta, USA \\
% ekhoury@pindrop.com}
}
\maketitle

\begin{abstract}
 Deep neural network based deepfake detection systems have achieved high levels of  accuracy on benchmark datasets and competitions. However, most models lack interpretability. It is challenging to extract reasoning from the network that can convince the human evaluator to trust the decision. Humans often rely on acoustic cues like unnatural pitch jitter, robotic intonation, acoustic artifacts, and unnatural sounding fricatives to judge the quality of the synthetic audio. This study explores the role played by the voiced and unvoiced regions of speech in discriminating synthetic from bonafide speech. A measure of signal periodicity is used to analyze speech into voiced and unvoiced components. Then, the graph attention based AASIST detection system is trained independently on each component.
This work compares the accuracy of deepfake detection system using voiced and unvoiced components and analyzes the results on the MLAAD dataset.
Our results show that unvoiced regions are particularly more effective in distinguishing synthetic (deepfake) speech from bonafide, and achieves an equal error rate of 6.62\%. When combined with voice regions through score-level fusion, the overall performance improves further, yielding a 5.82\% EER, a relative improvement of 49\% over the baseline system that uses the full audio.
%a relative improvement of almost 52\%  over the baseline system.

%Experiments reported in this paper show that unvoiced region helps to distinguish spoofed speech from bona fide and further enhance detection performance by combining with voiced region using the score-level fusion. 

% Deep neural network based deepfake detection systems have achieved high levels of accuracy on benchmark datasets
% and competitions. However, most models lack interpretability. It
% is challenging to extract reasoning from the network that can
% convince the human evaluator to trust the decision. Humans
% often rely on acoustic cues like unnatural pitch jitter, robotic
% intonation, acoustic artifacts, and unnatural sounding fricatives
% to judge the quality of the synthetic audio. This study explores
% the role played by the voiced and unvoiced regions of speech
% in discriminating synthetic from genuine speech. A measure of
% signal periodicity is used to analyze speech into voiced and
% unvoiced components. The AASIST detection system is then trained independently on each component. This work compares the detection accuracy of 
% voiced and unvoiced systems using the MLAAD dataset. Our results show that unvoiced region are particularly effective in distinguishing spoofed speech from bona fide,  with performance further improving when combined with voice regions using score-level fusion.

\label{sec:abstract}
\end{abstract}

\section{Introduction}
\label{sec:introduction}
% Show the challenge posed by deepfakes
% Why deepfake detection is important?
The advent of generative AI technologies has revolutionized the speech synthesis, making it possible to generate realistic speech contents that are indistinguishable from human speech~\cite{Mai2023}.
%
%opened the floodgates for synthesizing ultra-realistic speech which humans fail to tell apart from a real person's voice~\cite{Mai2023}.
Multi-speaker speech synthesis systems now clone anyone's voice using just a few seconds of enrollment audio~\cite{Arik2018}.
These technologies combined with the ubiquity of everyone's voice on various open internet based platforms makes everyone vulnerable to voice spoofing attacks.
This not only threatens personal security but also undermines the integrity of automatic speaker verification (ASV) system, making it crucial to develop robust spoofing countermeasure for secure ASV. The ability to detecting synthetic speech across various communication platforms is becoming really important to maintain the trust in voice communication over digital channels.\\

%an urgent need of the hour to preserve the trust in voice communication over digital channels.

% They also pose a threat to automatic speaker verification (ASV) systems, making the speech spoofing countermeasures an absolute essential in ASV-based transactions.
% Detecting synthetic speech across all modes of communication is an urgent need of the hour to preserve the trust in voice communication over digital channels. 

Synthetic speech detection has been an active area of research for over a decade~\cite{Wu2015,wang2024asvspoof5crowdsourcedspeech}.
The goal of this task to distinguish bonafide speech samples from text-to-speech and voice conversion samples.
%Various neural network architectures have been developed to detect synthetic speech~\cite{Khan2023, Nautsch2021}.
Ongoing research into novel neural network architectures and loss functions has significantly advanced the state-of-the-art across a wide range of synthetic systems~\cite{Khan2023,Nautsch2021,wang2024asvspoof5crowdsourcedspeech}, including in-the-wild conditions~\cite{muller22_interspeech}.
The current state-of-the-art systems process raw speech waveforms as input, feeding them through a deep convolutional neural network, followed by temporal pooling of features to predict whether the presented speech utterance is bonafide or synthetic~\cite{tak2021end, Wang2021, jung2022aasist, Chen2020}.\\

% What explanations do models provide to convince humans that a given audio is deepfake?

Explainable models that support their predictions with specific evidence from the input space can help build human trust in these black-box systems and may even assist forensic experts in identifying subtle signatures of synthetic speech.
Extracting interpretable explanations for a DNN model's predictions is a challenging task and is relatively a new area of research.
Authors in~\cite{lim2022detecting} studied several explainable artificial intelligence (XAI) methods used for image classification. Particularly, they looked at Deep Taylor~\cite{montavon2017explaining}, integrated gradients~\cite{Sundararajan2017AxiomaticAF}, and layer-wise relevance propagation~\cite{bharadhwaj2018layer}. Their empirical results concluded that those approaches tend to focus on the regions of unvoiced speech as well as the first three formants when voiced speech. 
Similarly, a recent study~\cite{Ge2022} that applied SHapley Adaptive exPlanations (SHAP)~\cite{Lundberg2017} to identify target regions in the speech spectrum found artefacts associated to particular regions in the voiced and unvoiced speech frames. \\

%% Humans are also capable of detecting deepfakes relying on artifacts in voiced and unvoiced regions.
%Humans often judge the quality of synthetic speech by focusing on specific artifacts.
%Perceptually, we all look for robotic intonations,  unnatural pitch jitter, fricatives and stop consonants sounding uniform and white-noise-like to assess the quality of a deepfake audio. 
%In this paper we ask the question whether the spoofing countermeasure systems be guided to focus on artifacts which
%
% The articulation of voiced and unvoiced sounds in speech are fundamentally controlled by the control of the vocalc chords and the transitions might hold clues that can tell the difference between genuine and synthetic speech. 
\begin{figure}[!t] 
    \centering
    \includegraphics[trim={0cm 0.25cm 0cm 0cm},clip,width=\linewidth]{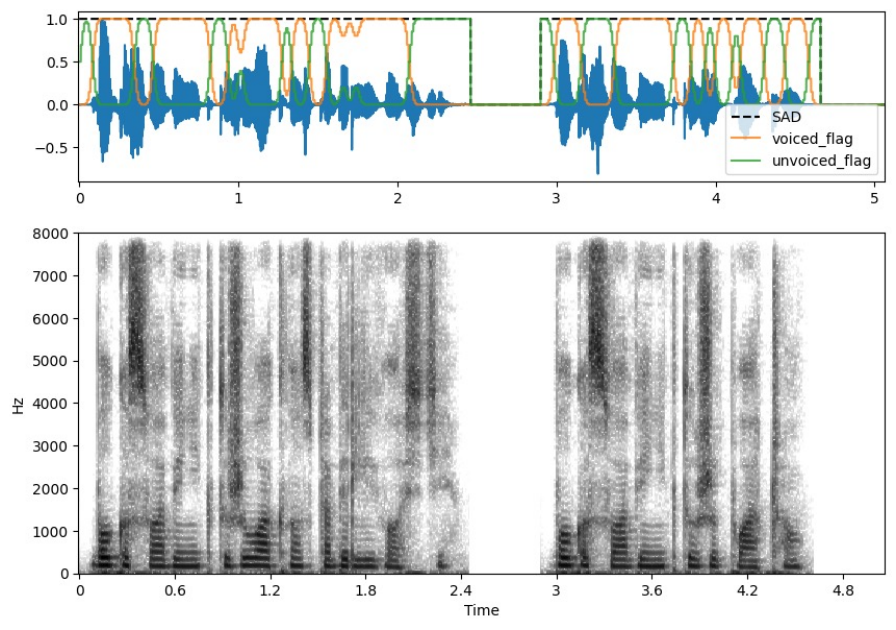} 
    \caption{Spectrogram of an example utterance from the MLAAD train set. Top panel shows the waveform along with the speech activity detection (SAD), voiced and unvoiced smoothed flags. The bottom panel shows the spectrogram of the utterance, highlighting its acoustic features.}
    \label{fig:voiced-unvoiced-flags}
\end{figure}

Speech production involves a complex coordination of several muscular movements. The vibration of the vocal cords play a central role in producing vowel sounds which form the major content of speech~\cite{stevens2000acoustic}.
But the vocal fold vibrations are intermittently stopped to produce unvoiced sounds like unvoiced stop consonants and fricatives.
Free flowing speech involves a rhythmic dance between the voiced and unvoiced states of vocal folds. 
Unvoiced fricative signals look similar to white noise but are often colored by the adjacent sounds due to co-articulation.
The extent of such co-articulatory effects in unvoiced sounds is not uniform in bonafide speech.
In contrast, text-to-speech synthesis and voice conversion systems often produce unnatural waveforms for the fricatives and stops.
Hence the unvoiced regions could hold important cues for discriminating bonafide from synthetic speech.\\

In the literature, phoneme specific models have been explored for detecting replay attacks with fair success.~\cite{Mochizuki2018} studied the impact of the phoneme-based pop-noise of the microphone, while~\cite{wang2021experimental} directly explored the use of voiced and unvoiced segments. They both achieve high accuracy on the ASVspoof 2017 dataset~\cite{kinnunen17_interspeech}.
Additionally, for synthetic speech detection, studies such as~\cite{muller2021speech,zhang122021effect,mari2022sound,zhang2023impact} highlighted the importance of silence regions for spoofing detection. Most of their work rely on the ASVspoof 2019 dataset~\cite{Todisco2019} that has silence at the beginning and end of the speech utterances, making it difficult to judge the significance of voice and unvoiced segments.\\

However to the best of our knowledge, we are the first to explore the use of voiced and unvoiced phones to detect synthetic speech. We conducted our experiments using the recently proposed multi-lingual audio anti-spoofing (MLAAD) dataset~\cite{muller2024mlaad}. Fig~\ref{fig:voiced-unvoiced-flags} illustrates an audio example from the MLAAD dataset, with both the waveform and the spectrogram. It also shows the voiced and unvoiced regions.
Our goal is to address the following research questions: 

% What is the objective of this paper?
%The goal of this work is to address the following research questions:
\begin{itemize}
    \item Can existing spoofing detection systems accurately detect synthetic speech by analyzing separately the voiced and unvoiced regions?
% \item What is the contribution of each component towards the overall accuracy?
   % \item What is the detection accuracy when anti-spoofing systems are trained solely on either unvoiced or voiced segments of speech? 
   \item Are the unvoiced or voiced segments complementary in improving detection performance?
\end{itemize}

% We conducted experiments using the multi-language audio anti-spoofing (MLAAD) dataset~\cite{muller2024mlaad} to address these specific research questions.  %Details about the dataset are presented in section~\ref{sec:datasets}. 
The remainder of this paper is organized as follows. Section~\ref{sec:methods} describes our methodology for segmenting speech utterances into voice and unvoiced components, and model architecture details. Our experiments and results are presented in Section~\ref{sec:experiments} followed by conclusion in Section~\ref{sec:conclusion}.
\\

% In this work, we propose a very simple method to temporally segment a speech utterance into voiced and unvoiced regions based on pitch detection.%All the files in the MLAAD dataset are split into voiced and unvoiced waveforms.
% We train spoofing detection system using the widely use graph neural network (GNN) based model, AASIST~\cite{jung2022aasist}.
% Section~\ref{sec:methods} outlines our methodology for segmenting speech utterances and provides additional details about the model architecture.
% % We train a countermeasure on each component (original, speech, voiced, and unvoiced) of the ASVSpoof dataset, perform evaluations and analyze the results as an attempt to answer the above questions.
% Our experiments and results are presented in Section~\ref{sec:experiments} followed by conclusions in Section~\ref{sec:conclusion}.

% What are the questions we are trying to answer?
% How we go about doing it?
% How is the paper organized?

\section{Voiced and unvoiced-based synthetic speech detection}
\label{sec:methods}
In this section, we explain the method used to analyze the speech signals by dividing them into voiced and unvoiced components. We also describe the model architecture used for the deepfake detection.
%\vspace{-0.25cm}
\subsection{Voiced and unvoiced segmentation}
\label{sec:segmentation}
Speech sounds can be categorized into two fundamental components: voiced and unvoiced phones. Voiced sounds occur when the vocal folds vibrate during the speech production, whereas unvoiced sounds do not. In the literature several algorithms have been proposed to decompose speech signal into these components~\cite{D'Alessandro1998,Deshmukh2005,Jackson2000,Yegnanarayana1998}. This decomposition has been applied for various purposes such as voice analysis, acoustic phonetics and speech recognition. % If there is vibration of the vocal folds during the production of a phone then it is defined as voiced. % Several algorithms have been proposed in the literature for decomposing the speech signal into voiced and unvoiced components
% \cite{D'Alessandro1998, Deshmukh2005,Jackson2000, Yegnanarayana1998}.%Such decomposition has been used for various purposes like voice analysis, acoustic phonetics and speech recognition. 
In our work, we focus on the temporal segmentation of speech into voiced and unvoiced regions using the pitch detection and estimation that can be used as a measure of voicing. There are several existing algorithms to detect and estimate the pitch~\cite{Slaney1990,Rabiner1977,deCheveigne2002,Rabiner1976}.
Most of them are based on peak finding on the auto-correlation function. 
In this paper, we use the probabilistic YIN (pYIN) algorithm to identify whether a given speech frame is voiced, using the implementation from the Librosa library~\cite{McFee2015}. We analyze the speech signal with a frame-length of $80$ ms and a hop-length of $10$ ms to estimate the pitch, voicing probability and voicing flag. The binary voicing-flag estimated by pYIN\footnote{\url{https://librosa.org/doc/main/generated/librosa.pyin.html}} is used to segment the speech signal. 
This approach is particularly effective when working with clean datasets where non-speech noise is minimal.\\ %\textcolor{blue}{Ganesh please write about VAD process here.}

To improve the segmentation process, we smoothed the voiced and unvoiced flags using a Hamming window with a width of $10$ frames. 
This smoothing helps to reduce the abrupt bursts in the speech signal due to discontinuities.
In order to exclude the non-speech silences from our analysis, we used the Web-RTC based speech activity detection (SAD) implementation\footnote{{\url{https://github.com/wiseman/py-webrtcvad}}} with a severity parameter of 2 to get the speech/non-speech binary mask.
The input audio signal $x$ is multiplied with the binary SAD mask to obtain the $x_{speech}$ signal.
\begin{equation}
    x_{speech} = SAD(x)*x, 
    \label{eq:flag}
\end{equation}

The $x_{speech}$ signal is multiplied with the smoothed voicing flag to generate the voiced component ($x_{voi}$), and is multiplied  with the unvoiced flag to generate the unvoiced speech ($x_{unv}$) as follows:
\vspace{-0.2cm}
\begin{equation}
    x_{voi} = voicing\_flag(x)*x_{speech}, 
     % \textit{voi} = \textit{SAD}*(\textit{voicing-flag}),
     \label{eq:voc_flag}
\end{equation}
\begin{equation}
    x_{unv} = (1- voicing\_flag(x))*x_{speech},
    \label{eq:unv_flag}
\end{equation}

Fig~\ref{fig:voiced-unvoiced-flags} shows the $SAD$, $voiced\_flag$, and $unvoiced\_flag$ for an example utterance from the MLAAD training subset.
%
% \textcolor{blue}{Where VAD is voice activity detection using webrtcvad tool\footnote{{\url{https://github.com/wiseman/py-webrtcvad}}}.}
%
Note that after smoothing, the voiced and unvoiced flags may overlap slightly at the boundaries, which helps include the transition regions and also minimizes the discontinuities in the speech signal.
In this work, we performed deepfake detection experiments using the full audio $x$, SAD segmented $x_{speech}$, voiced component $x_{voi}$, and unvoiced component $x_{unv}$. The same model architecture was used in all four scenarios. This architecture is described in the next section.

\subsection{Synthetic detection model}
\label{sec:system}
In the literature several Deep Neural Networks (DNNs) have been proposed to detect the synthetic speech.
Recent research has extensively explored large speech foundation models like Wav2vec2.0~\cite{baevski2020wav2vec}, Conformer~\cite{gulati2020conformer}, WavLM~\cite{chen2022wavlm} and Whisper~\cite{radford2023robust} to achieve the state-of-the-art performance. The main focus of our work is to highlight the importance of voiced and unvoiced regions in the speech for explainable audio deepfake detection, rather than achieving state-of-the-art performance.
Hence, we selected the widely-used AASIST~\cite{jung2022aasist} model. ASSIST is a graph attention based system that employs a RawNet2-based encoder\cite{tak2021end} to extract higher-level spectro-temporal features representation directly from the raw waveform. RawNet2 encoder consists of a bank of 70 sinc filters and six residual blocks to extract the spectral and temporal representations. A heterogeneous graph attention layers and max graph
operations are then used to model the local spectro-temporal dependency. 
Finally, the output scores are generated using a readout operation and a linear output layer.

\vspace{-0.2cm}
\section{Datasets}
\label{sec:datasets}
We performed our experiments using the MLAAD dataset~\cite{muller2024mlaad}. It consists of 52 different state-of-the-art synthetic speech generation systems~\cite{muller2024mlaad}. 
More details about the generation systems can be found in the Coqui-TTS~\cite{Eren_Coqui_TTS_2021} and HuggingFace repositories.
We manually labelled the acoustic models and vocoders based on the available metadata. %~\cite{klein2024source}.
For experiments, we divide the data into three disjoint partitions: train, development, and evaluation sets.
Bonafide samples were sourced from the multilingual M-AILABS dataset~\cite{mailabs}. The training and development partitions were created with a set of 9 different attacks (vits, xtts-v1, xtts-v2, tacotron2, tacotron2-dca, tacotron2-ddc, glow-tts and neural-hmm), whereas the
evaluation set was created with a set of 5 unseen attacks (bark, capacitron, fastpitch, overflow and tortoise-tts). 
Details of the data partitions are presented in Table~\ref{tab:MLAAD}.

\begin{table}[t]
 \setlength\tabcolsep{20pt}
\caption{MLAAD train, development and evaluation protocols for binary classification task.}
\vspace{-0.25cm}
% \centering
\resizebox{0.5\textwidth}{!}{
\begin{tabular}{|c|c|c|c|}
\hline
\multirow{2}{*}{\textbf{Types}} & \multicolumn{3}{c|}{\textbf{Partitions}} \\
\cline{2-4}
 & \multicolumn{1}{c|}{train}
 & \multicolumn{1}{c|}{dev}
 & \multicolumn{1}{c|}{eval} \\ \hline

Bonafide  & 33225 & 3656 & 4438 \\ \hline
Synthetic & 54867 & 6133 & 5000 \\ \hline

\end{tabular}
}
\label{tab:MLAAD}
\end{table}

% \resizebox{8.5cm\textwidth}{!}
% {
% \begin{tabular}{|c|c|c|c|}
% \hline
% \multirow{2}{*}{\textbf{Types}} & \multicolumn{3}{c|}{\textbf{Partitions}}  \\
% \cline{2-4}
%  &         \multicolumn{1}{c|}{train} & \multicolumn{1}{c|}{dev} & \multicolumn{1}{c|}{eval} \\ \hline
% \multicolumn{1}{|c|}{Bonafide}                 & \multicolumn{1}{c|}{33225}               & \multicolumn{1}{c|}{3656}                  & \multicolumn{1}{c|}{4438}            \\ \hline
% \multicolumn{1}{|c|}{Synthetic}                & \multicolumn{1}{c|}{54867}                & \multicolumn{1}{c|}{6133}                    & \multicolumn{1}{c|}{5000}             \\ \hline              
% \end{tabular}
% }
% \label{tab:MLAAD}
% \end{table}

\section{Experiments and results}
\label{sec:experiments}
The objective of the experiments in this paper is to assess the importance of voiced and unvoiced components of speech for deepfake detection. 
We trained four ASSIST models on each of the following input types - $full$ audio, SAD segmented $speech$, $voiced$, and $unvoiced$ components.
The model architecture was trained on each of these conditions with segments of approximately 4 seconds duration (64,600 samples). 
% AASIST model was trained on voice and unvoiced audio components with a segments of approximately 4 seconds duration (64,600 samples). 
% We also performed experiments using original audio (without voice and unvoiced segmentation), with only speech part after applying webrtc VAD and with \textit{voi} and \textit{unv} components. 
We followed the same training recipe and model configurations available in this repository\footnote{\url{https://github.com/clovaai/aasist/tree/main}}.
% We perform segmentation of the speech signal into voice and unvoiced components in on-the-fly fashion.
For training, we used the Adam optimiser with a mini-batch size of 16, a fixed learning rate of 0.0001, and trained for 30 epochs. The best model was selected based on the lowest equal error rate (EER) on the development set.
%All the results are reported in-terms of Equal-error-rate (EER).
% All models are evaluated with the Equal Error Rate (EER), and the tandem Detection Cost Function (t-DCF) metrics \cite{Kinnunen2018}.
% For the t-DCF metric we report the minimum achievable t-DCF (min t-DCF) value for each system by sweeping across the detection thresholds as described in \cite{Kinnunen2018}. 
\vspace{-0.1cm}
\subsection{Standalone results}
\begin{table*}[t]
\setlength\tabcolsep{12pt}
\center
\caption{ Pooled EER and attack-wise results in terms of FAR for the
MLAAD evaluation set on different scenarios.}
\vspace{-0.1cm}
\resizebox{0.99\textwidth}{!}
{
\begin{tabular}{|c|c|c|c|c|c|c|}
\hline
\multirow{2}{*}{\textbf{Scenario}}      & \textbf{Pooled} & \multicolumn{5}{c|}{\textbf{FAR  of attacks @ EER threshold}}                                                                                                                                                  \\ \cline{3-7}
& \textbf{EER}  & \textbf{bark} & \textbf{capacitron} & \textbf{fastpitch} & \textbf{overflow} & \textbf{tortoise tts} \\ \hline
\textbf{full-audio}        & 11.40\%                           & 10.90\%      & 6.30\%             & 0.00\%             & 0.00\%            & 39.80\%                                     \\ \hline
\textbf{speech-only}          & 10.10\%                           & 15.70\%       & 16.20\%             & 0.00\%             & 0.00\%            & 18.60\%                                     \\ \hline
\textbf{voiced}          & 12.26\%                           & 30.10\%       & 14.20\%            & 0.10\%             & 0.00\%            & 16.90\%                                     \\ \hline
\textbf{unvoiced}        & 6.62\%                            & 2.10\%        & 1.60\%              & 0.10\%            & 0.00\%            & 29.30\%                                     \\ \hline
\textbf{voiced+unvoiced} & 5.82\%                            & 7.80\%        & 3.10\%              & 0.10\%             & 0.00\%            & 18.10\%                                     \\ \hline
\end{tabular}
}
\label{tab:results}
\end{table*}

We first assess independently the performance of all four systems on the evaluation set of MLAAD as detailed in Table~\ref{tab:results}.
%Model trained on $voiced$ segments was evaluated on the $voiced$ segments of the utterances in the evaluation set, and similarly for the other conditions.
%Results for the four AASIST models is presented in the first column of 
The baseline system trained on full utterances achieved an EER of 11.40\%. The speech-only system achieved slightly better results with an EER of 10.10\%. Additionally, the voiced-only system got slightly worse EER of 12.26\%. But interestingly, the unvoiced-only system achieved an EER of 6.62\%, much lower than the three previous scenarios. 
A detailed breakdown of these results across attack types is also provided in columns 3-7 of Table~\ref{tab:results}. 
It shows the false acceptance rate (FAR) of each attack type at the threshold corresponding to the EER on the full evaluation set.
Based on this analysis, it is interesting to see that the unvoiced-only system works very well on the synthetic speech generated by Bark, capacitron, fastpitch and overflow, but it does not perform well on tortoise-tts samples. This is possibly because its mean spectrum is much closer to bonafide than the other synthetic systems.
\begin{figure}[!t] 
    \centering
    \includegraphics[trim={0cm 1.1cm 0cm 1.2cm},clip,width=0.5\linewidth]{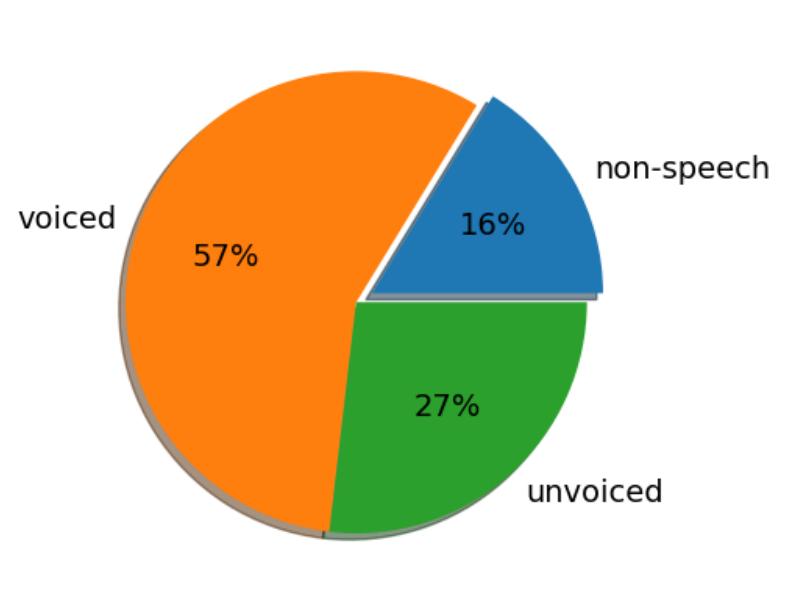}
    \caption{Proportion of utterance duration attributed to non-speech, voiced, and unvoiced regions in the train subset of MLAAD.}
    \label{fig:speech-voiced-unvoiced-pie}
\end{figure}

\subsection{Fusion results}
We use the linear logistic regression approach from scikit-learn to combine the scores from both voiced and unvoiced systems. We applied mean and std normalization on both voice and unvoiced development scores and trained logistic regression fusion model on them. The fusion results are shown in the last row of Table~\ref{tab:results}. 
The EER of the combined system is 5.82\%, which shows significant improvement of 52\% over the voiced system and 12\% over the unvoiced system. Our fusion results clearly show that both voice and unvoiced systems are complementary in improving deepfake detection performance. More importantly, the combined system yields 49\% reduction in EER (from 11.40\% to 5.82\%) compared to the standard technique that uses the full audio.  

\begin{figure}[!t] 
    \centering
    \includegraphics[trim={0cm 0.2cm 0cm 0.2cm},clip,width=1\linewidth]{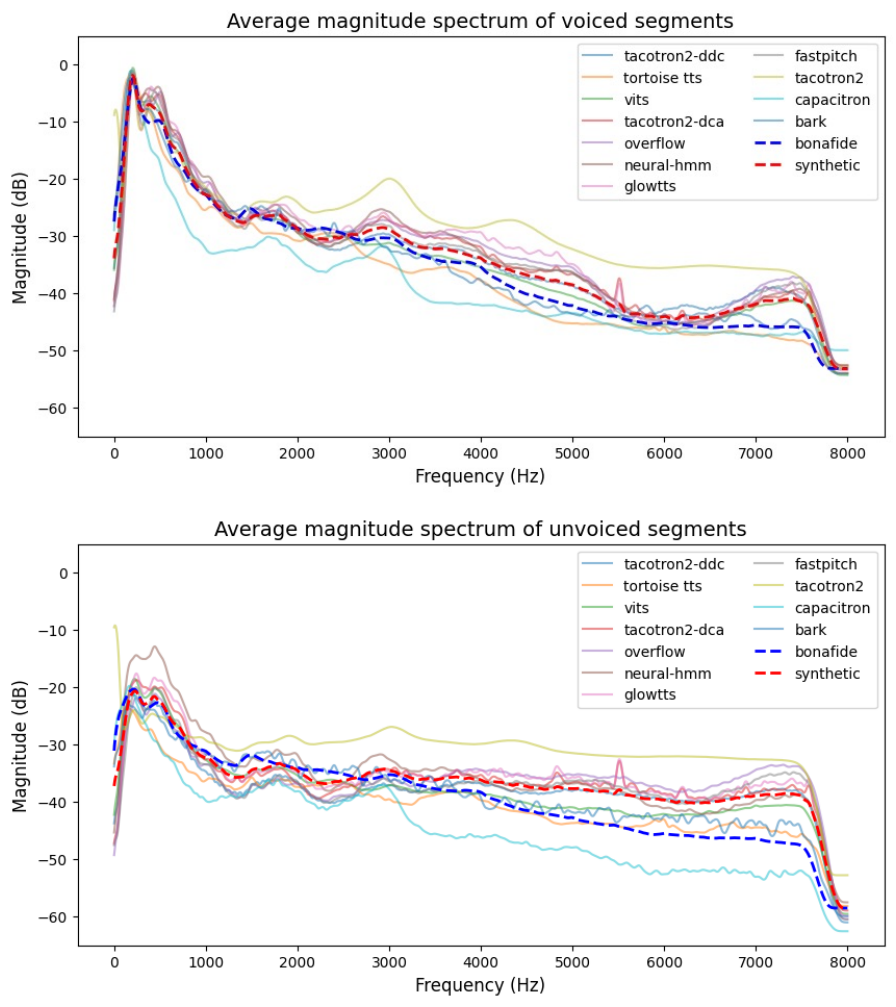} 
    \caption{Average spectrum of the voiced and unvoiced segments for bonafide and synthetic utterances from the MLAAD dataset.}
    \label{fig:avg-spectrum}
\end{figure}

\subsection{Analysis}
\begin{figure}[!t] 
    \centering
    \includegraphics[trim={0cm 0.18cm 0cm 0.2cm},clip,width=1\linewidth]{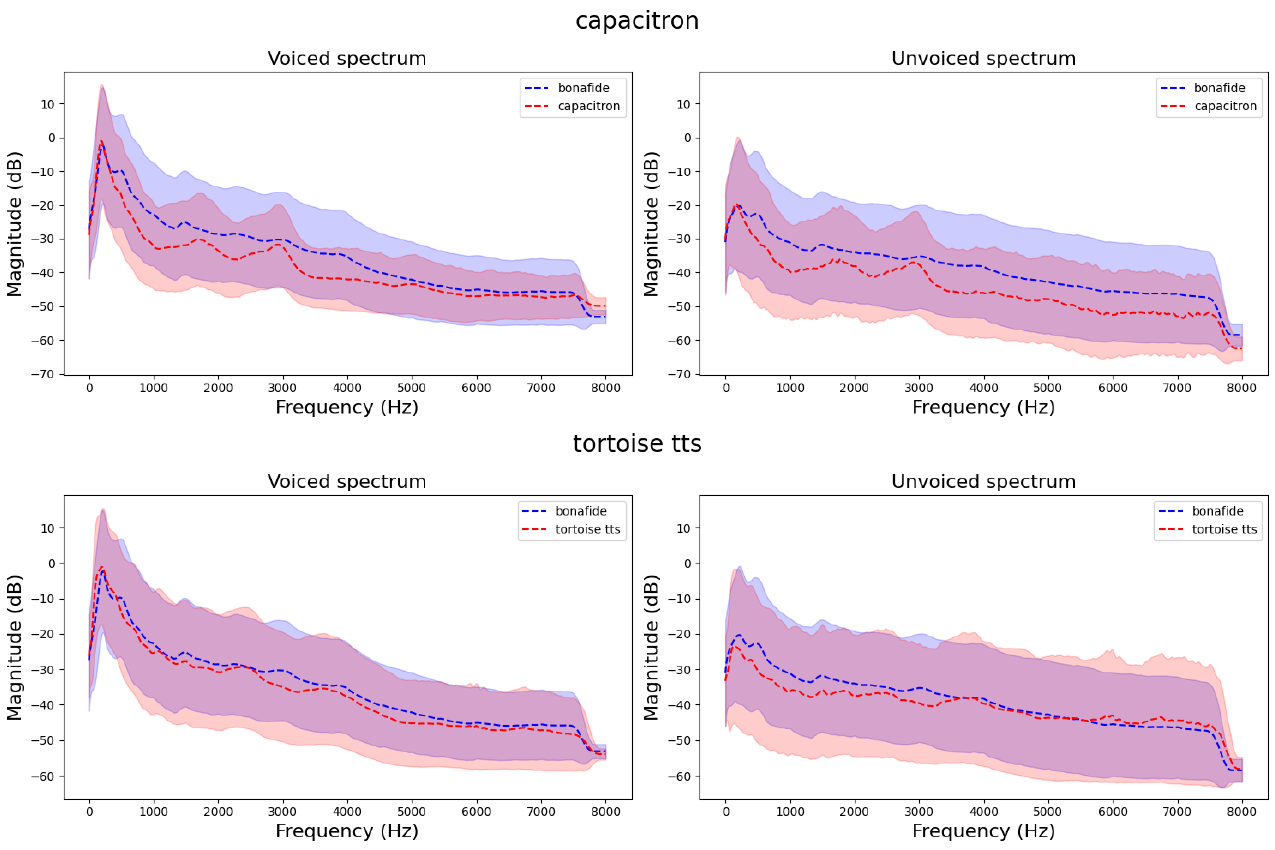} 
    \caption{Comparing the spectral mean and standard deviation of bonafide, fastpitch and tortoise-tts samples.}
    \label{fig:spectral-mean-std}
\end{figure}

The duration of an utterance consists of both non-speech and speech regions. The speech region in turn consists of both voiced and unvoiced regions.
Fig~\ref{fig:speech-voiced-unvoiced-pie} shows the pie chart of the proportion of an utterance attributed to non-speech, voiced and unvoiced regions.
The duration of the unvoiced segments of speech are on an average only 27\% of the audio duration. 
Yet, the unvoiced components outperform the voiced components for deepfake detection.
We investigated the spectrum of the signal to understand why the unvoiced regions of speech contain more discriminative features for deepfake detection.
Fig~\ref{fig:avg-spectrum} shows the average spectrum of bonafide and synthetic classes.  
We see that the average spectra of bonafide and synthetic utterances are more overlapping for the voiced segments compared to that of unvoiced segments.
For the unvoiced segments, there is a clear departure of the synthetic speech spectrum in the higher frequencies.
Fig~\ref{fig:spectral-mean-std} illustrates the spectral mean and standard deviations for Capacitron and Tortoise tts spoofing attack types. 
We observe that the spectra of the unvoiced segments are well separated in the higher frequencies for the Capacitron TTS spoofing attack.
However, the Tortoise-TTS model produces a spectrum much closer to bonafide speech which makes is harder to detect compared to the other attack types in the evaluation set.
Perhaps the TTS systems which are optimized for perceptual quality, tend to synthesize the lower frequency spectrum better than the higher frequencies.
The higher frequency spectral artefacts in the synthetic speech might go unnoticed to the human auditory system. We also observed that the FAR of the model on fastpitch, and overflow TTS was consistently close to 0\% across all the systems. 
The fastpicth and overflow models used in creating the MLAAD dataset were both single speaker TTS models trained on the LJSpeech dataset.
That might be the reason for the 100\% accuracy on these two attack types.

% \begin{figure}[!t] 
%     \centering
%     %\includegraphics[width=\linewidth]{contrib/average_spectrum_all} 
%     \includegraphics[trim={0cm 0cm 0cm 0.2cm},clip,width=\linewidth]{contrib/average_spectrum_all} 
%     \caption{Average spectrum of the voiced and unvoiced segments for bonafide and spoofed utterances from the MLAAD dataset.}
%     \label{fig:avg-spectrum}
% \end{figure}

\section{Conclusions}
\label{sec:conclusion}
In this paper we studied the role of voiced and unvoiced speech regions for synthetic speech detection.
Our experimental results, performed on the MLAAD dataset, show that the deepfake detection system trained on the unvoiced regions achieve better EERs compared to the systems trained on the full audio, speech-only regions or voiced regions. %Our experiments, performed on the MLAAD dataset, show that the unvoiced system outperforms voiced system and baseline systems by a substantial margin. Unvoiced system also outperform all other systems for 3 out of 5 spoofing attacks in the evaluation set. 
Fusion experiments also show that voice and unvoiced systems are complementary. The combined system yields 49\% relative reduction in EER over the baseline system that uses the full audio.
Future work will investigate an attention-based single end-to-end system that can possibly learn key features from both voiced and unvoiced regions.
% The results showed that the LFCC and LFB features provide an advantage to the unvoiced regions for detecting fake speech.
% Spectrogram and raw waveform representations are better for detecting deepfakes from the voiced regions.
% The score level fusion of the systems trained on voiced, unvoiced and the original baseline reduces the EER  from 3.19\% to 2.48\% on the official ASVspoof 2019 LA protocol with the LCNN architecture and LFCC frontend.
% We achieve 2.99 and 1.21 points of absolute reduction in EER on the systems trained on the LFB and spectrogram frontends. 
% The system fusion does not provide any improvement with the Rawnet-2 architecture.
% The analysis performed in this paper can help improve future model architectures towards efficiently capturing the artifacts from the voiced and unvoiced regions for detecting deepfakes. 

\bibliographystyle{IEEEtran}
\bibliography{mybib}

\end{document}